\documentclass[
    aps,
    prx,
    superscriptaddress,
    twocolumn,
    a4paper,
    floatfix,
    longbibliography
]{revtex4-2}

\usepackage{blindtext}
\usepackage{calc}
\usepackage{tikz}

\usepackage[american]{babel}
\usepackage[utf8]{inputenc}

\usepackage{grffile} % allow for period characters in filename

\usepackage{newtxtext,newtxmath}
\usepackage{microtype}

\usepackage{amsmath}
\usepackage{amssymb}
\usepackage{bbm}% bold math
\usepackage{mathtools}
\usepackage{dsfont}% einheitsoperator
\usepackage{braket}
\usepackage{cancel}
\usepackage{slashed}
\usepackage{float}
\usepackage[scr=boondoxo]{mathalpha}
\usepackage{subfig}

\usepackage{graphicx}% I Fig.~files

\usepackage{siunitx} %Fuer einheiten und nicefrac

\usepackage{ragged2e}
\usepackage{array}
\usepackage{tabularx}
\usepackage{booktabs}
\usepackage[export]{adjustbox}

\definecolor{cset-aps-blueberry}{RGB}{28,128,158}
\definecolor{cset-aps-blue}{RGB}{46,44,184}
\definecolor{cset-aps-turquoise}{RGB}{0,67,88}
\definecolor{cset-aps-limegreen}{RGB}{190,219,67}
\definecolor{cset-aps-green}{RGB}{31,138,112}
\definecolor{cset-aps-yellow}{RGB}{255,225,25}
\definecolor{cset-aps-orange}{RGB}{253,116,0}
\definecolor{cset-aps-red}{RGB}{219,0,43}

\makeatletter
\DeclareRobustCommand{\Arrow}[1][]{%
\check@mathfonts
\if\relax\detokenize{#1}\relax
\settowidth{\dimen@}{$\m@th\rightarrow$}%
\else
\setlength{\dimen@}{#1}%
\fi
\sbox\z@{\usefont{U}{lasy}{m}{n}\symbol{41}}%
\begin{picture}(\dimen@,\ht\z@)
\roundcap
\put(\dimexpr\dimen@-.7\wd\z@,0){\usebox\z@}
\put(0,\fontdimen22\textfont2){\line(1,0){\dimen@}}
\end{picture}%
}
\makeatother

\usepackage{hyperref}
\hypersetup{%
    colorlinks=true,
    linkcolor={cset-aps-red},
    linkbordercolor={cset-aps-red},
    filecolor={cset-aps-orange},
    filebordercolor={cset-aps-orange},
    citecolor={cset-aps-blue},
    citebordercolor={cset-aps-blue},
    urlcolor={cset-aps-green},
    urlbordercolor={cset-aps-green},
    menucolor={cset-aps-limegreen},
    menubordercolor={cset-aps-limegreen},
    breaklinks=true,
    pdfborderstyle={/S/U/W 2},
    pdfpagemode=UseOutlines,
    pdfstartpage={1},
}

\usepackage[noabbrev,capitalize]{cleveref}
\crefname{equation}{Eq.}{Eqs.}
\crefname{figure}{Fig.}{Figs.}
\crefname{table}{Tab.}{Tabs.}
\crefname{section}{Sec.}{Secs.}
\crefname{subsection}{Subsec.}{Subsecs.}
\crefname{subsubsection}{Subsubsec.}{Subsubsecs.}

\usepackage{lipsum}

\usepackage{placeins}
\usepackage{epsfig}
\usepackage[inline]{enumitem}
\setlist*[enumerate]{label=(\arabic*)}

\begin{document}

\title[Optimized Circuit Cutting for QAOA Sampling Tasks]{Optimized Circuit Cutting for QAOA Sampling Tasks}
\author{Friedrich Wagner}
\email{friedrich.wagner@iis.fraunhofer.de}
\address{Fraunhofer Institute for Integrated Circuits IIS, Nuremberg}
\author{Christian Ufrecht}
\address{Fraunhofer Institute for Integrated Circuits IIS, Nuremberg}
\author{Martin Braun}
\address{DATEV eG, Nuremberg}
\author{Daniel D. Scherer}
\address{Fraunhofer Institute for Integrated Circuits IIS, Nuremberg}
\date{July 2025}

\begin{abstract}
Circuit cutting was originally designed to retrieve the expectation value of an observable with respect to a large quantum circuit by executing smaller circuit fragments.
In this work, however, we demonstrate the application of circuit cutting to a pure sampling task.
In particular, we sample solutions to an optimization problem from a trained QAOA circuit.
Here, circuit cutting leads to a broadening and shift of the bitstring distribution towards suboptimal values compared to the uncut case.
To reduce this effect, we minimize the number of required cuts via integer programming methods.
On the other hand, cutting reduces the circuit size and thus the impact of noise.
Our experiments on quantum hardware reveal that, for large circuits, the effect of noise reduction outweighs the derogative effects on the bitstring distribution.
The study therefore provides evidence that circuit cutting combined with optimized cutting schemes
can both scale problem size and mitigate noise for near-term quantum optimization.
\end{abstract}

\maketitle

\section{Introduction}
This technical report provides a summary of experimental results from quantum computing hardware for the technique introduced in our previous work \cite{Herzog2024}.
In Ref.~\cite{Herzog2024}, we propose the combination of a classical size reduction method for quadratic unconstrained binary optimization (QUBO), introduced in~\cite{Wagner_2025},
with circuit cutting for sampling from the quantum approximate optimization algorithm (QAOA).
Tailoring the classical reduction method to the requirements of circuit cutting allows us to increase the size of the QUBO instances that can be handled by currently available quantum hardware.
In this report, we showcase the feasibility of the approach by proof-of-principle experiments on actual quantum hardware.
To this end, we consider the problem of finding a maximum cut in a graph (MaxCut), which is a special case of QUBO, and
apply our approach to MaxCut instances on graphs of up to 25 nodes.
The results demonstrate a significant noise mitigation through the application of circuit cutting. 

Circuit cutting is a technique to decompose a quantum circuit into smaller subcircuits that can be run individually on quantum hardware.
This procedure, in principle, allows to run large quantum circuits on small quantum hardware.
Circuit cutting can be roughly categorized into gate cutting \cite{Hofmann2009,Mitarai2021,Ufrecht2023,Piveteau2022_circuitcut,Schmitt2024,Harrow2024},
where quantum gates are cut, and wire cutting \cite{Peng2020, Lowe2022, Brenner2023, Harada2023},
where qubit wires are cut.
In our approach, we make use of the wire cutting paradigm and leverage the fact that
the minimum number of wire cuts needed per QAOA layer to create independent subcircuits
is given by the cardinality of a minimum vertex separator in the QUBO graph \cite{Lowe2022}.

Circuit cutting introduces a sampling overhead that increases exponentially with the number of cuts
and quickly becomes prohibitive if circuit cutting is performed naively.
To keep the overhead manageable, we initially undertake the following classical preprocessing steps: 
We first identify a minimum balanced vertex separator,
that is, we pinpoint two sets of nodes of approximately equal size that are interconnected by a low-cardinality vertex separator.
Finding such a separator is known to be NP-hard~\cite{BUI1992}.
In practice, integer programming techniques allow to find solutions to instances with hundreds of nodes in the order of seconds~\cite{Althoby2020}.
Having identified a minimum balanced vertex separator, we perform \emph{graph shrinking} \cite{Wagner_2025, fischer2024quantumclassicalcorrelationsshrinking},
a technique to reduce the number of nodes in a QUBO problem while keeping track of node correlations
such that a solution to the original problem can be reconstructed from a solution to the reduced problem.
Via graph shrinking, we further reduce the size of the vertex separator and thus the number of required wire cuts.
Subsequently, we apply wire cutting to the QAOA circuits corresponding to the shrunk graph, utilizing methods proposed by Peng et al.~\cite{Peng2020}
and Harada et al.~\cite{Harada2023} in an alternating manner across subsequent QAOA layers.
This strategy not only reduces the sampling overhead compared to using only the method of Peng et al.~but
also ensures that only one-way classical communication is required.
This allows to run all circuits sequentially and obtain the solution of the original problem
in the postprocessing step without the need of establishing real-time classical communication links between quantum computers.
This is crucial for our experiments, since real-time connected quantum computers are still at an experimental stage and currently not available for public users.

Circuit cutting is an exact method for estimating expectation values of observables.
However, given the values of the variational parameters in QAOA circuits, retrieving actual solutions to the optimization problem reduces to a pure sampling task.
Nonetheless, we propose the application of circuit cutting to this sampling task by reconstructing bitstrings corresponding to the uncut circuit~\cite{Herzog2024}.
While the reconstructed bitstring distribution is broadened and shifted compared to the uncut distribution,
theoretical bounds ensure that any bitstring of the original problem can be found with high probability~\cite{Lowe2022}
when the sampling budget is increased by a constant factor with respect to the number of qubits~\cite{Herzog2024}. 

We emphasize that the approach \cite{Herzog2024} is only a first step into the direction of leveraging circuit cutting for combinatorial optimization.
Indeed, we could perform ``classical cutting'' by simply iterating over all possible assignments of values to the nodes in the vertex separator.
For each assignment of values, we run QAOA on the two resulting independent subgraphs.
This approach would incur lower sampling overhead than quantum circuit cutting.
In particular, the sampling overhead would not increase with the number of QAOA layers. 
Thus, more refined circuit cutting techniques are needed that exploit the structure of QAOA circuits
in order to beat this classical overhead.
Finally, it should be noted that all QUBO instances discussed in this report can be solved to proven optimality 
in fractions of a second using state-of-the-art classical solvers on a standard laptop. 

\section{Results}
In this section, we provide computational results supporting our claim that circuit cutting can
indeed mitigate noise in sampling tasks on currently available quantum devices.
The key insight is that the noise reduction due to smaller circuits can over-compensate the distribution broadening incured by circuit cutting.

\paragraph{Instances.}
We consider four MaxCut instances with 10, 15, 20 and 25 vertices.
These instances resemble toy examples for the identification of key customers in a social network,
a real-world application of MaxCut~\cite{best_practise_2025}.
The instance data is summarized in Table~\ref{tab:instances}.

\paragraph{Procedure.}
For each instance, we apply the following procedure.
First, we compute a low-cardinality, balanced vertex-separator by solving the integer program introduced in~\cite{Herzog2024}.
Subsequently we shrink the vertex separator to a single vertex using the linear programming approach introduced in~\cite{Wagner_2025}.
Then, we construct the corresponding two-layer QAOA circuit with pre-computed parameters.
The parameters are obtained using a local optimizer and a noiseless simulation of the circuits.
We initialize the local optimizer with parameters corresponding to a quantum annealing schedule as proposed in~\cite{Sack_2021}.
Subsequently, we apply wire cutting to the qubit corresponding to the single separating vertex.
In order to obtain two independent circuit partitions, we need to apply two cuts, one in each layer of the QAOA circuit.
The first cut is performed using the method of Harada et al.~\cite{Harada2023}.
For the second cut, we apply the method of Peng et al.~\cite{Peng2020}.
We then sample from the two circuit fragments and reverse the entire decomposition procedure to ultimately retrieve a
sequence of solutions to the original MaxCut instance.
\begin{table}
    \begin{tabular}{cccc}
         $|V|$& $|E|$ & $|S|$ & $c^*$  \\
         \midrule
         10  &  13  &  2  &  11 \\
         15  &  20  &  3  &  18 \\
         20  &  29  &  4  &  26 \\
         25  &  38  &  3  &  33 \\
    \end{tabular}
    \caption{Instance data. For each instance, we give the number of nodes $|V|$, the number of edges $|E|$,
    the size of the separator $|S|$ and the value of an optimum MaxCut solution $c^*$.}
    \label{tab:instances}
\end{table}
For comparison, we also sample from the uncut two-layer QAOA circuit corresponding to the shrunk MaxCut instance.
This allows us to quantify the effect of circuit cutting on noisy devices.
For both approaches (cut and uncut) we sample from
a real quantum device and a noiseless simulator.

\paragraph{Setup.}
For solving integer and linear programs, we use the open-source solver SCIP~\cite{BolusaniEtal2024OO}.
For circuit construction, backend communication and noiseless simulation, we use the open-source software Qiskit~\cite{javadiabhari2024quantumcomputingqiskit}.
We employ the local optimizer COBYLA~\cite{Powell1994ADS} for parameter pre-computation.
The quantum device in our experiments is \emph{ibm\_aachen}~\cite{IBMQuantum}.
For each instance and method, we take $100,000$ samples.
Theoretical bounds \cite{Herzog2024} guarantee that the probability
to retrieve each bitstring is suppressed by at most a factor of
$1/\kappa=1/12$ compared to the uncut setup.
This can be compensated for by increasing the sample budget by a factor of $\kappa$. 

\begin{figure}
\includegraphics[width = 1 \linewidth, trim = {0 0 0 1.45cm}, clip]{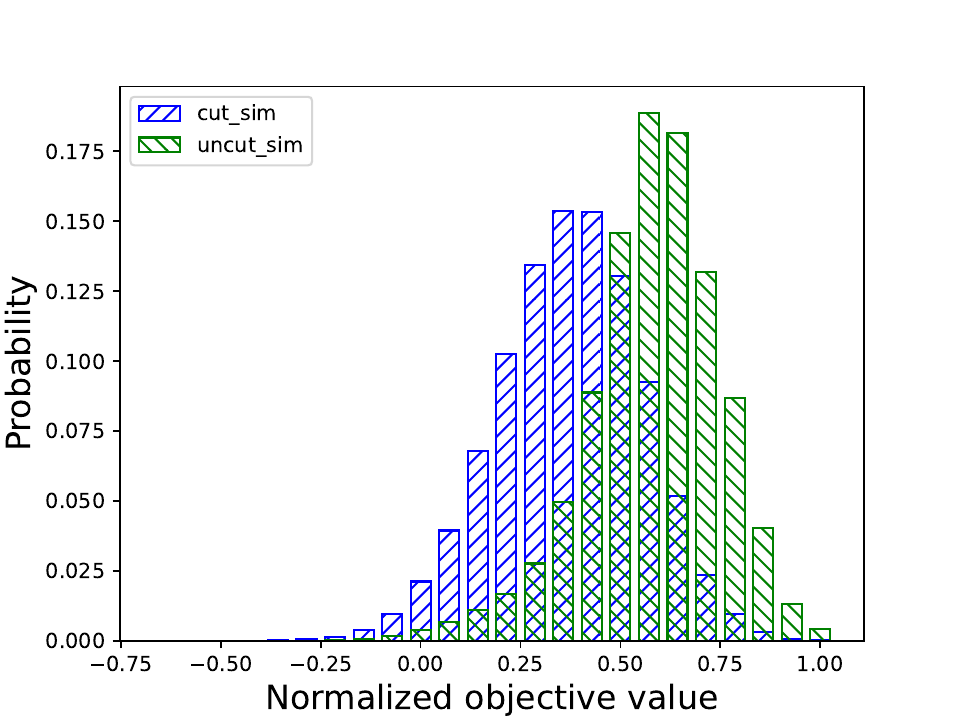}
\caption{Histograms retrieved from cut (//) and uncut (\textbackslash\textbackslash) circuits executed on a noise-free simulator for the MaxCut instance on 25 nodes.}
\label{fig:hist_25_sim}
\end{figure}

\paragraph{Results.}
A single run of our procedure returns a sequence of solutions to the considered MaxCut instance $G=(V,E)$.
This sequence can be translated into a histogram of objective values, see Figure~\ref{fig:hist_25_sim} for an example.
To allow for a comparison of objectives across different MaxCut instances, we use the normalized objective
\begin{align}
    r=\frac{c-c_0}{c^*-c_0} \leq 1
\end{align}
where $c_0=|E|/2$ is the expected objective when sampling solutions uniformly at random
and $c^*$ is the optimum objective.
Thus, a value of $r=0$ corresponds to the objective value of random sampling while
a value of $r=1$ corresponds to an optimum solution.

In Figure~\ref{fig:hist_25_sim}, we compare the histograms
retrieved from the cut and uncut circuits
for the 25-node MaxCut instance using a noise-free simulator.
As expected, the distribution from the cut circuit
is broadened and shifted to {suboptimal values compared to the uncut distribution.
This is due to the fact that we employ circuit cutting
to sample bitstrings from the uncut circuit instead of estimating expectation values~\cite{Herzog2024}.
In Figure~\ref{fig:hist_25_ibm}, we compare the histograms
for the cut and uncut circuits using actual hardware.
\begin{figure}
\includegraphics[width = 1  \linewidth, trim = {0 0 0 1.45cm}, clip]{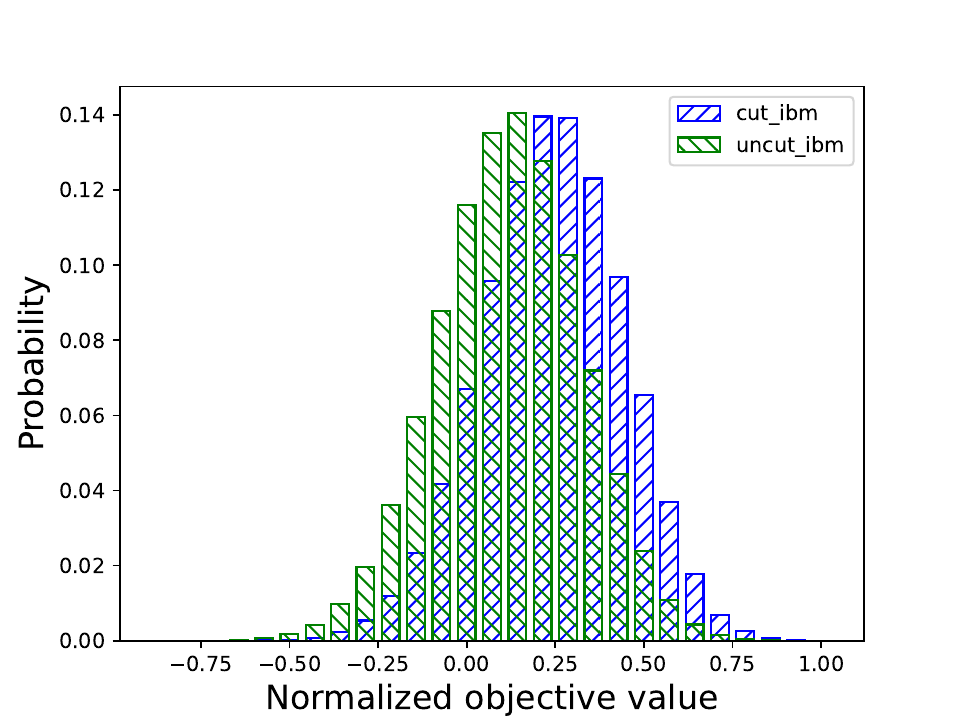}
\caption{Histograms retrieved from cut (//) and uncut (\textbackslash\textbackslash) circuits executed on a noisy device for the MaxCut instance on 25 nodes.}
\label{fig:hist_25_ibm}
\end{figure}
Interestingly, in this case, the cut distribution is 
in fact shifted closer to the optimum compared to the uncut distribution.
Cutting reduces the circuit width and depth
and thus decreases the influence of noise and errors.
In our test case, this effect over-compensates the shift and broadening of the distribution incurred by using circuit cutting for sampling tasks.
For the other three MaxCut instances with 10, 15 and 20 nodes, we
did not observe this improvement of the cut distribution.
This can be seen in Figure~\ref{fig:quantils} where we
compare the 95th percentiles of the sampled distributions.  
\begin{figure}
\includegraphics[width = 1  \linewidth, trim = {0 0 0 1.3cm}, clip]{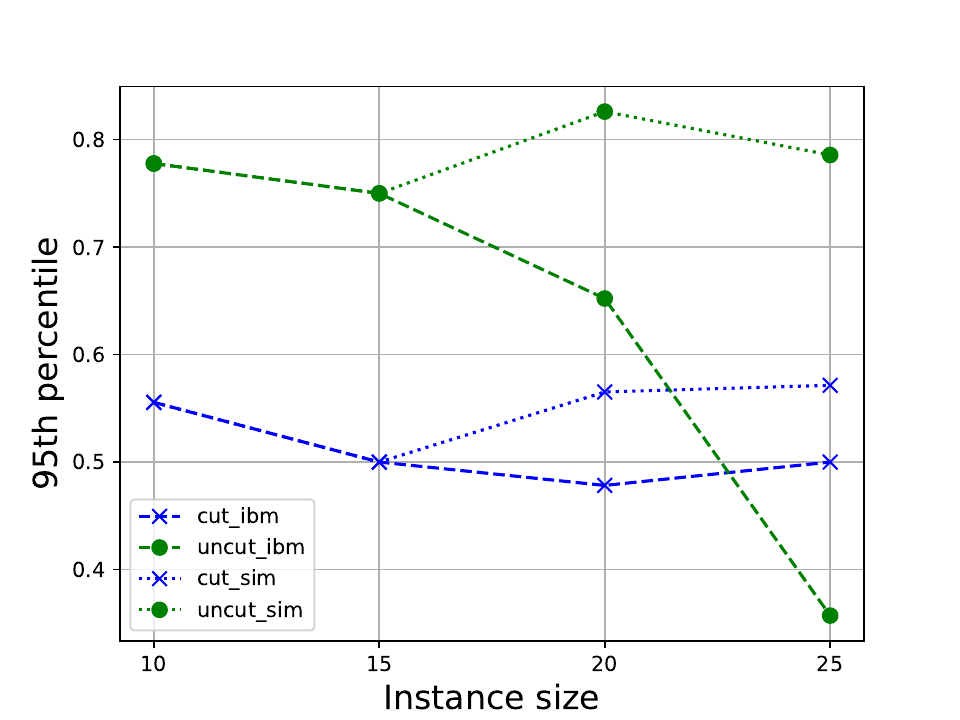}
\caption{95th percentiles for the distributions retrieved from cut (crosses) and uncut (circles) circuits on an ideal simulator (dotted) and a real device (dashed) in dependence of the instance size.}
\label{fig:quantils}
\end{figure}
We attribute this observation to the fact that the influence of noise
typically increases exponentially with the circuit size
and the noise reduction due to cutting is thus more effective for larger circuits.

\section{Conclusion}
In our experiments, we applied circuit cutting together with classical pre-processing based on integer programming
to the task of sampling bitstrings from trained QAOA circuits.
For the 25-node graph, the noise mitigation due to the reduced width and depth of the cut circuits
even improved the quality of the bitstring distribution compared to an uncut execution.
For smaller graphs, this benefit vanishes, indicating that the payoff grows with circuit size and device noise. 
These proof-of-principle experiments suggest that circuit cutting can in principle
act as an effective noise-mitigation tool for combinatorial optimization.
However, the sampling overhead required to guarantee that high-quality solutions are found quickly becomes
prohibitive as the number of layers increases.
Moreover, the simplest form of ``classical cutting'' incurs lower sampling overhead than circuit cutting.
Practical applications therefore require refined, low sampling-overhead circuit-cutting techniques tailored to combinatorial optimization via QAOA.

\begin{acknowledgements}
The authors thank L.~S.~Herzog and M. Periyasamy for helpful discussions.
We acknowledge the use of IBM Quantum services
for this work. 
The research was funded by the project
QuaST, supported by the Federal Ministry for Economic Affairs and Climate Action on the basis of a
decision by the German Bundestag.
\end{acknowledgements}

\bibliography{CuttingBib, CVRP, optimization}

\end{document}